\documentclass{sig-alternate-05-2015}
\pdfoutput=1

\usepackage{xspace}
\usepackage[svgnames,rgb,usenames,dvipsnames,svgnames,table]{xcolor}
\usepackage{subfigure}
\usepackage{graphicx}
\usepackage{epsfig}
\usepackage{epstopdf}
\usepackage{array}

\begin{document}

\title{STEPS: Predicting place attributes \\ via spatio-temporal analysis}

\author{
\alignauthor
Shuxin Nie, Abhimanyu Das, Evgeniy Gabrilovich\titlenote{Corresponding author}\hspace{-1mm},\\
\vspace{1mm}
Wei-Lwun Lu, Boris Mazniker, Chris Schilling\\
\vspace{2mm}
\affaddr{Google, 1600 Amphitheatre Parkway, Mountain View, CA 94043}
\email{\{shnie|abhidas|gabr|weilwunlu|mazniker|cschill\}@google.com}
}

\maketitle

\newcommand{\attrib}{attribute\xspace}
\newcommand{\attribs}{attributes\xspace}
\newcommand{\Attrib}{Attribute\xspace}
\newcommand{\Attribs}{Attributes\xspace}

\newcommand{\method}{\textsc{STEPS}\xspace}
\newcommand{\methodSeq}{\mbox{\textsc{STEPS}}\xspace}
\newcommand{\methodEmb}{\mbox{\textsc{STEPS-E}}\xspace}

\newcommand{\todo}[2]{{\color{red} TODO! {\sc #1}
: \textit{#2}}}

\newcolumntype{R}{@{\extracolsep{3mm}}r@{\extracolsep{0pt}}}%

\begin{abstract}
In recent years, a vast amount of research has been conducted on learning people's interests from their
actions. Yet their \emph{collective} actions also allow us to learn something about the world, in particular, infer attributes of places people visit or interact with.
Imagine classifying whether a hotel has a gym or a swimming pool without ever talking to the people who stayed there. Or imagine predicting whether a restaurant has a happy hour or a romantic atmosphere without
asking its patrons. Algorithms we present can do just that.

Many web applications rely on knowing \attribs of places, for instance, whether a particular restaurant has WiFi or offers outdoor seating. Such data can be used to support a range of user experiences, from explicit query-driven search to proactively recommending places the user might like. However, obtaining these \attribs is generally difficult, with existing approaches relying on crowdsourcing or parsing online reviews, both of which are noisy, biased, and have limited coverage. Here we present a novel approach to classifying place \attribs, which learns from patrons' visit patterns based on anonymous observational data.

Our method, \method, learns from aggregated sequences of place visits. For example, if many people visit the restaurant on a Saturday evening, coming from a luxury hotel or theater, and stay for a long time, then this restaurant
is more likely to have a romantic atmosphere. On the other hand, if most people visit the restaurant on weekdays, coming from work or a grocery store, then the restaurant is less likely to be romantic. We show that such transition features are highly predictive of place \attribs.
We also introduce a variant of our method, \methodEmb, which builds a high-dimensional embedding model trained on co-visitation data, and completely eliminates the need for feature engineering.
%
%
In an extensive empirical evaluation, \method nearly doubled the coverage of a state of the art approach thanks to learning from observational location data, which allowed our method to reason about many more places.

\end{abstract}

\section{Introduction}
\label{sec:intro}

\vspace{-2mm}

\begin{quote}
{\em ``You know my method. It is founded upon the observation of trifles.''}

-- Sherlock Holmes.
\end{quote}

\vspace{-2mm}


In recent years, numerous web services and applications have been built to facilitate access to information about brick-and-mortar businesses and places such as restaurants, hotels, parks, or tourist sites.
These include review sites and recommendation apps (TripAdvisor, Yelp, Zagat), online business directories (Urbanspoon, Yellowpages), mapping services (Google Maps), travel sites (Hotels.com, Expedia), and even web search. Many of these services rely on knowing \attribs of places to help users narrow down their search.
For example, recommendation web sites often categorize restaurants using a variety of \attribs, such as whether a restaurant is a fine-dining or casual place, whether it offers free WiFi or outdoor seating, whether it takes reservations or offers take-out.
Similarly, travel sites  allow users to search for hotels based on \attribs such as whether the hotel is frequented by business customers or leisure travellers, and whether it has amenities such as a gym or swimming pool. Therefore, identifying these \attribs is a critical component in many user-facing applications.

However, obtaining these \attribs of places in a scalable manner is challenging. Existing approaches
have traditionally taken one of two routes: (a) Crowdsourcing the task,
or (b) Inferring \attribs from text analysis of online reviews about the place. The former approach explicitly asks visitors
to manually specify \attribs of the place. This approach is hard to scale to a large number of places and their \attribs, because people are often unwilling to spend time answering questions without any perceived near-term gain. Furthermore, it is difficult to ensure high quality of crowdsourced labels. More recently, several studies have focused on mining customer reviews
to extract key \attribs of a place, along with customer opinions about these \attribs~\cite{liubook}. While this approach does not require manual labor (beyond developing a classifier), it also suffers from a number of weaknesses. Reviews are often noisy, verbose and ambiguous, and it is  difficult to reliably parse review text to extract information about specific \attribs. Reviews also have a major coverage problem. New places have few or no reviews, and even for older places the review text often covers only a few main \attribs. For example, many restaurant reviews talk about the cuisine type, prices or ambience, but only a few (if at all) may explicitly mention whether the restaurant requires reservations, offers take-out, or has outdoor seating (see Table~\ref{tab:main-results-rest} for the long list of \attribs we predict). Finally, reviews are biased because only a handful of people choose to write them, and usually they were either extremely happy or unhappy with the service.

We take a fundamentally different approach, and study the movement of people to infer \attribs of places they visit. As a motivating example, imagine you stumbled upon a coffee shop you have never visited before. Over two dozen people are waiting in line. The coffee must be good, you say to yourself, because all those people waiting patiently must know what they are doing. At that very moment you have inferred an attribute of a physical place by observing how people act around it. In this paper, we present algorithms that do just that.


We introduce \method, \underline{S}pacio-\underline{TE}mporal analysis of \underline{P}lace attribute\underline{S}, which learns from trajectory patterns of patrons' visits to the place.
\method derives spatial and temporal features from anonymous aggregated Location History data that our users have proactively chosen
to share with us.\footnote{Users can switch it off at any time via My Account. This is essentially the same data we use in Google Now
to notify users about the best time to visit their favorite museum.~\cite{GoogleEuropeBlog}}
The data comes in the form of aggregated visit sequences (before/after visiting a given place), along with place categories (e.g., restaurant or grocery store), binned arrival time, and duration. Similar data is available in the form of place check-ins on Foursquare, Facebook or Yelp, or geo-coded tweets on Twitter. However, the anonymous visit data has much higher coverage (as it does not require explicit check-in) and temporal resolution (we can compute visit durations and arrival times). It is also much less biased than reviews because no effort is required from the visitor. In this paper, we show how this micro-scale visit data allows us to learn new macro-scale facts about the world (place \attribs).

\method uses features based on sequences of other places visited by patrons of a given place, which are often predictive of place \attribs. For example, if many people visit a particular restaurant soon after visiting a park or a beach, then the restaurant is more likely to have outdoor seating because it appeals to patrons who like to be outdoors. If a restaurant is often visited for short periods of time in the afternoon, after visiting another restaurant and before going home, then it is likely to offer good desserts. On the other hand, if many people go to a restaurant directly from home or work and stay there for a long time, it is probably good for meals rather than desserts.
(See Section~\ref{sec:qualitative} for additional examples of visit patterns we use as prediction signals.)
One might think that many of these patterns are only weakly correlated with the \attrib we are trying to predict.
A key insight of this paper is that when we aggregate multiple weak signals over a large population,
we can classify many place \attribs with high accuracy and coverage.

We formulate the problem of \attrib classification in a supervised learning setting, and use manually labeled training data.
\method employs a semi-automatic approach to feature generation, which constructs a large number of features based on several simple (hand-chosen) properties of co-visited places, for example, their categories (e.g., grocery store or park) and the durations of visits. We then use these features to train a binary classifier for each \attrib.

We also introduce a  variant of \method called \methodEmb, which completely automates feature construction and learn features directly from the data.
\methodEmb performs collaborative filtering on the co-visitation patterns of people and places to create a high-dimensional
\underline{e}mbedding (hence ``\textsc{-E}'') of each place in an underlying latent space. The dimensions of the latent space are then treated as automatically-constructed features for
training binary classifiers for each \attrib.
\methodEmb uses low-rank matrix factorization~\cite{koren}
to automatically create $1000$-dimensional feature vectors for each place. The latent features in \methodEmb are much fewer
than the fine-grained features created in \method. However, as evident from the experiments in Section~\ref{sec:eval}, they are rich enough to capture
many of the spatio-temporal visit patterns useful for \attrib classification.


%

The contributions of this paper are threefold. First, we proposed a novel approach to classifying \attribs of a place based on other places frequently visited by its patrons.
Second, we employed the richness of aggregated anonymous location data to automatically construct features that characterize nearby locations visited and the time spent there. It should be emphasized that this data is entirely observational, hence no extra effort is required to collect it. Third, we performed a comprehensive evaluation of our approach using real world data spanning several dozen different \attribs of thousands of places (restaurants and hotels). We compared our results to a state of the art baseline that mines the text of online reviews to infer \attribs. We showed that our approach significantly increases the coverage of the baseline method, while exhibiting competitive classification accuracy. Since reviews are not always available and may not mention all the relevant \attribs, relying on reviews is a major limitation, which our proposed approach alleviates and nearly doubles the classifier coverage. We also note that the nature of the location data is orthogonal to that of the review text.
As a result, when we combine features from the two data sources (reviews + place visits), we observe a further improvement
in classification performance. Finally, we present a qualitative study that explains in depth how our method works on several \attribs.

\section{Background}
\label{sec:background}

With the rapid increase in location-aware applications,
there has been a large body of work on location data mining and its various use-cases. Several papers have focused on collecting and modeling location data for
personalized Point-of-Interest (POI) recommendation. Ye et al.~\cite{ye}  modeled users' check-in behavior in
location based social networks, and proposed a collaborative POI recommendation algorithm using Naive Bayes methods.
Lian et al.~\cite{lian} proposed a matrix-factorization approach for POI recommendation, by jointly modeling users' geographic preferences and their
place-visit data from location based social networks. Zheng et al.~\cite{zheng2} mined a large scale GPS dataset to extract
correlations between user locations for personalized recommendations. Zheng et al.~\cite{zheng} used GPS data
along with user text comments to create a location-activity matrix for collaborative location and activity recommendations.
Cheng et al.~\cite{ccheng} considered temporal
and sequence information to predict a user's next-POI  by incorporating personalized Markov Chain information during matrix factorization.

Another direction in location data mining deals with modeling user trajectories and visit distribution patterns to predict a user's next destination or activity.
Li et al.~\cite{li} predicted user trajectories using
a probabilistic motion model trained on anonymized GPS-snippets. Cheng et al.~\cite{cheng} used check-in data from location based social networks, and learnt a
Hidden Markov Model to predict a user's activity and location at the next step. Lichman et al.~\cite{lichman} modelled a
user's spatial distribution using a mixture of Kernel Density Estimates. Kirmse et al.~\cite{kirmse} used location histories obtained from GPS and WiFi signals
to infer users' frequently visit places and commute patterns. Laio et al.~\cite{liao} modeled raw GPS traces using relational Markov networks to simultaneously infer
a user's significant locations and activities.

Somewhat related to our work is the recent paper by Zhong et al~\cite{zhong} who used location check-in records to create spatial and temporal location features of
users for predicting various user demographics such as age, gender, educational background, etc. However, the focus of our paper is different from \cite{zhong} (and
indeed, from most of the above papers on inference using location data) in that we model location data not to predict user attributes, but rather place attributes,
and hence our location features are aggregate, population level features, instead of per-user features.

While we are not aware of prior work on using location data for extracting place attributes, there has been a large body of work on inferring attributes of
products, places and other entities from text mining of online content. In particular, there exists a rich line of work in the area of opinion mining~\cite{liubook}, which performs
text analysis of documents, customer reviews or online discussion forums to identify attributes of an entity and extract customer opinions and sentiments for
these attributes. Several of these papers use a notion of frequent noun phrases to identify important attributes from reviews~\cite{ku, hu1, hu2, popescu, long}.
Another class of opinion mining papers involves leveraging relationships between potential attributes and sentiment-bearing
words (usually adjectives) to discover
relevant attributes, and their sentiment scores. Hu et al.~\cite{hu1} extracted low-frequency attributes by identifying the nearest noun phrases to
sentiment-bearing words. Qui et al.~\cite{qiu} used a double propogation method to jointly extract attributes and sentiment words. Zhuang et al.~\cite{zhuang}
mined relationships between attributes and sentiment words using a dependency grammer graph, and extracted valid attribute-sentiment pairs for movie reviews.
Jakob et al.~\cite{jakob} proposed a supervised CRF-based sequence modeling approach to extract attributes, using features such as the part of speech tags, string tokens,
dependency-parse-tree distances, and distance to sentiment words.
Wang et al.~\cite{wang} addresssed the problem of teasing out users' latent ratings on different topical attributes of hotel reviews
from users' overall review scores and their review text.


\section{Method}
\label{sec:method}

In this section, we provide a detailed description of the \method method, which learns from anonymous aggregated observational
data about people's place visits. \method constructs spatio-temporal features by mining historical data about sequences of place visits corresponding to a large group of people. Each place visit instance corresponds to a single visit by an anonymous person to a place, and also includes the category of the place (e.g., restaurant, hotel, or park), and discretized arrival time and visit duration. We use a taxonomy of several thousand place categories.

The key intuition here is that the sequence of places people visited before and after a given place, can reveal patterns that are correlated with \attribs of the place. Admittedly, this correlation might be quite weak for any individual pattern. However, our experimental results suggest that aggregating such patterns over a large population of place visitors leads to surprisingly strong prediction of place \attribs.

We construct a large set of spatio-temporal features using aggregated data for an entire population of visitors to a place.
The features can be logically grouped as follows.

\noindent $\bullet$ \textbf{Duration features:} These features capture the distribution of time people stay at a place. Intuitively, such features are useful for predicting \attribs of restaurants such as availability of WiFi, or whether it has fast service. For example, patrons may stay longer in a cafe with WiFi available, and they may stay less in a fast-food restaurant.

\noindent $\bullet$ \textbf{Arrival time features:} These features capture the distribution of peoples' arrival times to a place. Specifically, we create a set of real-valued features for each hour of the week, where the feature value corresponds to the fraction of people that visit the place at that hour. There are several \attribs for which this temporal signal is useful. For example, if a restaurant is frequented by patrons in the afternoon, it is more likely to offer good desserts. Similarly, if a restaurant is popular on Sunday mornings, it is likely that it serves a good brunch menu.

\noindent $\bullet$ \textbf{Occupancy features:} These features capture the occupancy distribution of a place (measured in terms of the fraction of people who visit the place at different hours of the week). For example, a restaurant offering brunch menu may be crowded between breakfast and lunch hours on Sundays. Note that while the arrival and duration features characterize people's visits, the occupancy features characterize the busyness of a place. Occupancy features essentially define the occupancy histogram of a place, by the hour of the week.

\noindent $\bullet$ \textbf{Transition features (previous visit)} characterize places visited \emph{before} the current place. We compute the distribution of peoples' visits to other places
within a time window ($1$, $4$, $8$, $16$ and $24$ hours). To ensure that such features generalize well, this distribution is computed not over the specific places people visit, but over place categories (e.g., restaurant, grocery store, hotel).
The value of each feature reflects the fraction of people who visited that place category in a given time window \emph{before} the target place visit (e.g., 2\% of visitors to the given restaurant have visited a grocery store in the previous hour). Several place \attribs can be predicted better using such features. For example, if a  cafe is frequented by people who visit another restaurant beforehand, then this cafe is more likely to offer good desserts. If a restaurant is popular among patrons who visit parks or beaches prior to it, then the restaurant is likely to have outdoor seating (as it appeals to people who like to be outdoors).

\noindent $\bullet$ \textbf{Transition features (next visit):} Analogous to the previ\-ous-visit features described above, we also compute the distribution of people's visits (to other places) \emph{after} visiting a given place, within a particular time window.
For example, if many people visit surf shops after a hotel, then the hotel is likely to have beach access. If many people visit a cafe in the morning after staying at a hotel, then it is likely that the hotel does not provide breakfast.

The above groups of features are instantiated for different place categories and time intervals, hence the total number of features we generate for each place is quite large (about $\sim$ 100K). This process is semi-automatic, because the only manual part is compiling the lexicon of place categories and the choice of time intervals. Cf.\ Section~\ref{sec:qualitative} for examples of features we use for predicting different \attribs.

We use supervised learning to train models for predicting individual place \attribs. We have a multi-labeled setting whereas each restaurant (or hotel) has multiple \attribs, and we predict each one independently by training a dedicated binary classifier. Note that our feature set is comprehensive enough to be predictive of all \attribs, and is not specific to any particular \attrib. We perform per-attribute feature selection (using a standard Mutual Information-based method), and retain up to 10,000 features per \attrib. All the attributes we predict are binary, hence we use binary classification models (however, \method is applicable to predicting multi-class \attribs too).

Specifically, for each \attrib we collect a ground-truth set of positive and negative labels corresponding to places that have this \attrib and places that do not have it. The labels are obtained from third-party aggregator sites as well as via crowdsourcing, as described in Section~\ref{sec:eval:datasets}.
In the experiments described in Section~\ref{sec:eval}, we used linear Support Vector Machines for classification, however, our \method method can work with any  binary classifier.


\subsection*{Automating feature generation with embeddings}

The \method method described in the previous section relies on human engineered features. We now propose a variant of \method called \methodEmb, which performs automated feature generation for \attrib classification, with only a small penalty in accuracy compared to using manual feature engineering. At a high level, \methodEmb generates a feature vector for each place using collaborative filtering on the (anonymous) person-place visit data.
We use low-rank matrix factorization~\cite{koren} on the person-place co-visit matrix (normalized to factor out location bias~\cite{lian}, as described below) to compute an embedding vector for each place in a latent low-dimensional space.
The dimensions of this vector are then used as features for representing the place.

Specifically, we first construct a person-place matrix $\mathbf{L}$ with
rows representing people and
columns representing pla\-ces. Note that the data in this matrix is completely anonymous. Every cell in the matrix contains
a boolean value representing whether the person has visited the place or not, and
has an associated weight corresponding to the number of times the person visits the place, capped by a maximum threshold. The weight represents a confidence score, and
captures the relative contribution of the cell to the matrix-factorization objective function. A well-known issue in geographical matrix factorization (see~\cite{lian}) is the tendency of people to have location bias. For example, people are more likely to visit restaurants in a small set of locations they are familiar with, and less likely to visit restaurants in unfamiliar areas. Thus, not visiting a restaurant in a familiar area should carry a stronger negative signal, and visiting a restaurant in new areas should carry stronger positive reinforcement.
Several approaches have been proposed to correct this location bias~\cite{lian}. We used a simple yet effective normalization heuristic, which divides the weight of each cell in the matrix (counting the person's visits to a place) by the number of other places this person has visited in a $2~km$ radius. Given this weighted co-visit matrix, we use standard low-rank matrix factorization to compute a low dimensional person embedding~$\mathbf{U}$ and place embedding~$\mathbf{V}$, by optimizing the following loss function (using Weighted Alternate Least Squares~\cite{koren}):
	$$\text{min}_{U, V} \sum_{i,j} W_{ij} \left( L_{ij} - U_i^T V_j \right)^2 + \lambda (\sum_i || U_i ||_2 + \sum_j || V_j||_2),$$
where $\mathbf{W}$ is the weighting matrix, and $\lambda$ is the L2-norm regularization constant.

Each place embedding is then used as a feature vector of the place to train binary classifiers for various \attribs, as described in the previous section. Since we use low-rank matrix factorization, the number of embedding dimensions is small enough that we do not need to use feature selection.

\section{Empirical evaluation}
\label{sec:eval}

We describe the datasets and the baseline, and then report the performance of our methods.
We also present ablation studies that explore the
utility of the different feature groups. All the results were obtained via 10-fold cross-validation. We evaluated the performance of place \attrib predictions using the area under the ROC curve (AUC).

\subsection{Datasets}
\label{sec:eval:datasets}

We evaluated our methodology on 2~large datasets with labeled data about \attribs of restaurants and hotels.

The restaurant dataset included 29 restaurant \attribs (all binary) listed in Table~\ref{tab:rest-attrib-list}. Admittedly, these \attribs are somewhat subjective (e.g., whether the restaurant is inexpensive), and are partly overlapping (e.g., cozy / quiet / romantic atmosphere). However, from a user perspective, these attributes are deemed useful for making restaurant recommendations, and we had ample human-labeled data for them, hence we used them to evaluate our method. Labeled examples corresponded to actual restaurants that had or did not have a given \attrib (positive / negative, respectively). The human labels were obtained from third-party aggregator sites, such that each label was confirmed by at least two sites (there were no label contradictions, resulting in essentially 100\% inter-rater agreement). Owing to lack of space, we do not show the exact numbers of examples for each \attrib, but the average number of positive examples per \attrib was 34K (median 24K) and the average number of negative examples was 54K (median 34K). These examples represent a sample of restaurants across several countries.

\begin{table}[ht]
\centering
\begin{tabular}{|l|l|l|}
  \hline
\textbf{\Attrib} & \textbf{Attrib type} & \textbf{Description}\\
  \hline
Wine & Notable & Good for wine\\
Takeout & Meal type & \mbox{Food takeout available\hspace{-1mm}}\\
Brunch & Meal type & Serves brunch\\
Happy hour & Meal type & Has happy hour\\
Upscale & Formality & Formal attire \\
Hip & Crowd & Attracts hip crowd\\
WiFi & Features & WiFi available \\
Romantic & Atmosphere & Romantic atmosphere \\
\mbox{Outdoor seating\hspace{-1mm}} & Features & Has outdoor seating \\
Breakfast & Meal type & Serves breakfast\\
Lunch & Meal type & Serves lunch\\
Dinner & Meal type & Serves dinner\\
Food & Intent & \mbox{Mainly visited for food\hspace{-1mm}}\\
     &        & (as opposed to drinks)\\
Drink & Intent & Main visit: drinks\\
Low price & Price & Inexpensive\\
Cozy & Atmosphere & Cozy atmosphere \\
Lively & Atmosphere & Lively atmosphere \\
Quiet & Atmosphere & Quiet atmosphere \\
Groups & Company & Good for groups \\
\mbox{No reservations\hspace{-1mm}} & Ease of entry & Reserv. not required \\
Usually a wait & Ease of entry & Longer wait time\\
Live music & \mbox{Entertainment\hspace{-1mm}} & Has live music\\
Fast food & Restaur. type & Serves fast food\\
Delivery & Features & Delivers food \\
Casual & Formality & Casual attire\\
Dessert & Notable & Good for desserts \\
Tea & Notable & Good for tea\\
Healthy & Food type & Serves healthy food \\
Vegetarian & Food type & Good veg. selection\\
   \hline
\end{tabular}
\caption{Restaurant \attribs.}
\label{tab:rest-attrib-list}
\end{table}
%

The second dataset included 16 binary hotel \attribs such as whether the hotel has a golf course, airport shuttle, beach access, free breakfast, laundry service, fitness center (the full list of attributes is given in the first column of Table~\ref{tab:main-results-hotels}).
Labeled examples corresponded to hotels that had or did not have a given \attrib. We obtained the labels from an in-house crowdsourcing project, using at least two raters per hotel.
The average number of positive examples per \attrib was 68K (median 65K) and the average number of negative examples was 91K (median 87K). These examples represent a sample of hotels across several countries.

\subsection{Baseline}
\label{sec:eval:baseline}

We compared the performance of our method, \method, with that of a baseline classifier trained on a corpus of text reviews. The reviews were crawled from public Google+ web pages for the respective restaurants and hotels. We pooled together all the reviews posted for each business, and represented each training instance with a bag of words. We did not use stemming, but removed stop words as well as words that occurred in reviews of only one place.
The average number of reviews per restaurant was 23.2 (median 5), and the average review length was 65.1 words (median 43). The average number of reviews per hotel was 112.4 (median 16), and the average review length was 74.8 words (median 59). We used linear SVM to build the baseline classifier. The performance of the baseline classifier is reported in the second column of Table~\ref{tab:main-results-rest} (restaurants) and Table~\ref{tab:main-results-hotels} (hotels).

\begin{table*}
\centering
\begin{tabular}{|l||c||c|c||c|c||@{}>{\hspace*{2mm}}c|c|c|c|c|}
  \cline{1-6} \cline{8-11}

\textbf{\Attrib} & \textbf{\small Reviews} & \textbf{\small \methodSeq} &  \textbf{Gain} & \textbf{\small Reviews}      & \textbf{Gain} & & \multicolumn{4}{c|}{\textbf{\small Ablation study:}}\\
    &            &            &          & \mbox{\hspace{-1.5mm}\textbf{\small + \methodSeq}\hspace{-1.5mm}} &   & &
    \multicolumn{4}{c|}{\textbf{\small Individual \method feature groups}}\\
    &            &            &          &           &         & &
    \mbox{\hspace{-1mm}\textbf{\small Duration}\hspace{-1mm}} & \mbox{\hspace{-1mm}\textbf{\small Arrival}\hspace{-1mm}} & \mbox{\hspace{-1mm}\textbf{\small Occupancy}\hspace{-1mm}} & \mbox{\hspace{-1mm}\textbf{\small Transition}\hspace{-1mm}}\\
  \cline{1-6} \cline{8-11}

Wine & 0.941 & 0.945 & 0.4\% & 0.968 & 2.9\% & & 0.876 & 0.841 & 0.846 & 0.909 \\
Takeout & 0.831 & 0.916 & 10.2\% & 0.929 & 11.8\% & & 0.842 & 0.773 & 0.801 & 0.893 \\
Brunch & 0.875 & 0.887 & 1.4\% & 0.921 & 5.3\% & & 0.771 & 0.8 & 0.776 & 0.827 \\
Happy hour & 0.928 & 0.891 & -4\% & 0.943 & 1.6\% & & 0.774 & 0.83 & 0.81 & 0.847 \\
Upscale & 0.916 & 0.95 & 3.7\% & 0.963 & 5.1\% & & 0.826 & 0.835 & 0.815 & 0.898 \\
Hip & 0.921 & 0.869 & -5.6\% & 0.939 & 2\% & & 0.747 & 0.728 & 0.724 & 0.852 \\
WiFi & 0.812 & 0.923 & 13.7\% & 0.931 & 14.7\% & & 0.86 & 0.823 & 0.834 & 0.905 \\
Romantic & 0.79 & 0.891 & 12.8\% & 0.892 & 12.9\% & & 0.816 & 0.759 & 0.785 & 0.855 \\
Outdoor seating & 0.846 & 0.845 & -0.1\% & 0.912 & 7.8\% & & 0.725 & 0.69 & 0.697 & 0.832 \\
Breakfast & 0.909 & 0.961 & 5.7\% & 0.972 & 6.9\% & & 0.889 & 0.958 & 0.948 & 0.921 \\
Lunch & 0.849 & 0.906 & 6.7\% & 0.926 & 9.1\% & & 0.829 & 0.863 & 0.863 & 0.832 \\
Dinner & 0.904 & 0.953 & 5.4\% & 0.967 & 7\% & & 0.87 & 0.918 & 0.916 & 0.899 \\
Food & 0.884 & 0.928 & 5\% & 0.946 & 7\% & & 0.81 & 0.86 & 0.84 & 0.886 \\
Drink & 0.916 & 0.946 & 3.3\% & 0.967 & 5.6\% & & 0.838 & 0.863 & 0.853 & 0.923 \\
Low price & 0.968 & 0.969 & 0.1\% & 0.992 & 2.5\% & & 0.951 & 0.896 & 0.918 & 0.965 \\
Cozy & 0.75 & 0.805 & 7.3\% & 0.84 & 12\% & & 0.7 & 0.678 & 0.684 & 0.777 \\
Lively & 0.737 & 0.733 & -0.5\% & 0.796 & 8\% & & 0.632 & 0.644 & 0.643 & 0.702 \\
Quiet & 0.755 & 0.817 & 8.2\% & 0.855 & 13.2\% & & 0.696 & 0.722 & 0.715 & 0.793 \\
Groups & 0.728 & 0.817 & 12.2\% & 0.841 & 15.5\% & & 0.757 & 0.737 & 0.747 & 0.791 \\
No reservations & 0.917 & 0.946 & 3.2\% & 0.968 & 5.6\% & & 0.918 & 0.861 & 0.887 & 0.915 \\
Usually a wait & 0.702 & 0.734 & 4.6\% & 0.798 & 13.7\% & & 0.623 & 0.629 & 0.621 & 0.691 \\
Live music & 0.723 & 0.863 & 19.4\% & 0.902 & 24.8\% & & 0.639 & 0.714 & 0.696 & 0.798 \\
Fast food & 0.918 & 0.945 & 2.9\% & 0.976 & 6.3\% & & 0.945 & 0.856 & 0.851 & 0.962 \\
Delivery & 0.891 & 0.888 & -0.3\% & 0.93 & 4.4\% & & 0.726 & 0.717 & 0.712 & 0.856 \\
Casual & 0.844 & 0.839 & -0.6\% & 0.89 & 5.5\% & & 0.648 & 0.628 & 0.587 & 0.803 \\
Dessert & 0.818 & 0.912 & 11.5\% & 0.942 & 15.2\% & & 0.892 & 0.867 & 0.854 & 0.907 \\
Tea & 0.896 & 0.938 & 4.7\% & 0.959 & 7\% & & 0.891 & 0.897 & 0.904 & 0.917 \\
Healthy & 0.764 & 0.801 & 4.8\% & 0.837 & 9.6\% & & 0.644 & 0.662 & 0.648 & 0.751 \\
Vegetarian & 0.852 & 0.861 & 1.1\% & 0.914 & 7.3\% & & 0.661 & 0.681 & 0.662 & 0.787 \\
    \cline{1-6} \cline{8-11}
Macro-average & 0.848 & 0.885 & 4.4\% & 0.918 & 8.3\% & & 0.786 & 0.784 & 0.781 & 0.852 \\

    \cline{1-6} \cline{8-11}
\end{tabular}
\caption{Classifying restaurant \attribs. Performance is reported as Area Under the Curve (AUC). Note that \methodSeq increases the coverage by 95.8\% compared to only using Reviews (cf.\ Section~\ref{sec:eval:main}).}
\label{tab:main-results-rest}
\end{table*}

\begin{table*}
\centering
\begin{tabular}{|l||c||c|c||c|c||@{}>{\hspace*{2mm}}c|c|c|c|c|}
  \cline{1-6} \cline{8-11}

\textbf{\Attrib} & \textbf{\small Reviews} & \textbf{\small \methodSeq} &  \textbf{Gain} & \textbf{\small Reviews}      & \textbf{Gain} & & \multicolumn{4}{c|}{\textbf{\small Ablation study:}}\\
    &            &            &          & \mbox{\hspace{-1.5mm}\textbf{\small + \methodSeq}\hspace{-1.5mm}} &   & &
    \multicolumn{4}{c|}{\textbf{\small Individual \method feature groups}}\\
    &            &            &          &           &         & &
    \mbox{\hspace{-1mm}\textbf{\small Duration}\hspace{-1mm}} & \mbox{\hspace{-1mm}\textbf{\small Arrival}\hspace{-1mm}} & \mbox{\hspace{-1mm}\textbf{\small Occupancy}\hspace{-1mm}} & \mbox{\hspace{-1mm}\textbf{\small Transition}\hspace{-1mm}}\\


  \cline{1-6} \cline{8-11}

Air conditioned & 0.953 & 0.934 & -2\% & 0.956 & 0.3\% & & 0.813 & 0.802 & 0.816 & 0.932 \\
All inclusive & 0.702 & 0.735 & 4.7\% & 0.756 & 7.7\% & & 0.583 & 0.606 & 0.632 & 0.719 \\
Golf coures & 0.782 & 0.821 & 5\% & 0.848 & 8.4\% & & 0.632 & 0.657 & 0.634 & 0.812 \\
Airport shuttle & 0.788 & 0.79 & 0.3\% & 0.82 & 4.1\% & & 0.6 & 0.666 & 0.661 & 0.788 \\
Bar or lounge & 0.919 & 0.873 & -5\% & 0.92 & 0.1\% & & 0.752 & 0.745 & 0.771 & 0.847 \\
Beach access & 0.919 & 0.916 & -0.3\% & 0.935 & 1.7\% & & 0.829 & 0.805 & 0.824 & 0.911 \\
Business center & 0.869 & 0.855 & -1.6\% & 0.879 & 1.2\% & & 0.718 & 0.736 & 0.751 & 0.842 \\
Fitness center & 0.949 & 0.899 & -5.3\% & 0.954 & 0.5\% & & 0.718 & 0.716 & 0.728 & 0.861 \\
Free breakfast & 0.892 & 0.839 & -5.9\% & 0.896 & 0.4\% & & 0.612 & 0.616 & 0.603 & 0.823 \\
Hot tub available & 0.895 & 0.83 & -7.3\% & 0.9 & 0.6\% & & 0.685 & 0.698 & 0.718 & 0.817 \\
Laundry service & 0.824 & 0.795 & -3.5\% & 0.829 & 0.6\% & & 0.645 & 0.656 & 0.663 & 0.785 \\
Restaurant & 0.947 & 0.912 & -3.7\% & 0.95 & 0.3\% & & 0.781 & 0.777 & 0.814 & 0.88 \\
Room service & 0.867 & 0.831 & -4.2\% & 0.871 & 0.5\% & & 0.706 & 0.697 & 0.713 & 0.809 \\
Spa available & 0.915 & 0.846 & -7.5\% & 0.92 & 0.5\% & & 0.741 & 0.73 & 0.743 & 0.816 \\
Pets allowed & 0.843 & 0.818 & -3\% & 0.864 & 2.5\% & & 0.639 & 0.633 & 0.669 & 0.813 \\
Smoke free & 0.778 & 0.773 & -0.6\% & 0.79 & 1.5\% & & 0.652 & 0.668 & 0.681 & 0.774 \\
  \cline{1-6} \cline{8-11}
Macro-average & 0.865 & 0.842 & -2.7\% & 0.881 & 1.8\% & & 0.694 & 0.701 & 0.714 & 0.827\\

  \cline{1-6} \cline{8-11}
\end{tabular}
\caption{Classifying hotel \attribs. Performance is reported as Area Under the Curve (AUC). Note that \methodSeq increases the coverage by 94.8\% compared to only using Reviews (cf.\ Section~\ref{sec:eval:main}).}
\label{tab:main-results-hotels}
\end{table*}

\subsection{Main findings}
\label{sec:eval:main}

We built the \methodSeq model using linear SVM over the features introduced in Section~\ref{sec:method}. In addition to linear SVM, we also experimented with a logistic regression classifier for both \methodSeq and the baseline, and the results were substantially similar (for both restaurants and hotels), hence we do not show them here owing to lack of space.

Table~\ref{tab:main-results-rest} shows the performance of the review baseline, our \method method, and the combination of both, for the restaurant dataset (cf.\ Section~\ref{sec:eval:datasets}). Each row in the table shows classifier performance for one \attrib, and the summary row reports overall performance macro-averaged over all the \attribs
(i.e., averaging the AUC values from the individual rows). We computed the gain in the summary row by comparing the macro-averaged AUC values for the method and the baseline.

As we can readily see, \methodSeq shows superior results on average, and demonstrates slightly inferior performance for only 6 out of 29 \attribs.

However, the true importance of these results stems from the dramatically \textbf{increased classification coverage} due to our method. Of all the restaurants in our database, reviews are only available for 30.4\% of restaurants, but \method features\footnote{To maintain the anonymous aggregated nature of the data, we computed \method features only for restaurants (hotels) that have been visited by at least 10~people.} can be computed for 59.6\% of restaurants --- \textbf{a relative coverage gain of 95.8\%} ! This happens because online reviews are not available for all businesses (especially for the newer ones), and even when they are available, they might not mention the \attribs we are interested in. People often write reviews to express a strong opinion (positive or negative) about the restaurant, but do not mention all the numerous applicable \attribs of the business. Therefore, the coverage of the review-based (baseline) classifier is limited. However, simply using observational data about how people get in and out of the restaurant, allows us to substantially increase the classification coverage, and reliably predict \attribs for almost twice as many restaurants.

As we see in Table~\ref{tab:main-results-rest} (column Reviews + \method), combining the review-based and the \methodSeq features leads to superior classification performance for all the \attribs.

We assessed the statistical significance of our results using the standard $t$-test. For all 23~\attribs where \method performance is higher than the baseline, the improvement is significant ($p < 0.01$). For the remaining 6~\attribs,
only 4 losses are significant (except Delivery and Casual), and all the losses are small in absolute terms. Importantly, the improvements shown by Reviews+\method are significant for all \attribs ($p < 0.01$).


Similarly, Table~\ref{tab:main-results-hotels} shows the classification performance of our method for hotel attributes (due to space constraints, we do not report individual,
per-attribute results). We compare the performance of \method with that of the review baseline, as well as report the performance of their combination. Here, \method shows slightly inferior results on average (with improvements for 3 out of 16 \attribs). However, this is more than compensated by \textbf{substantial improvements in coverage}. Of all the hotels in our database, reviews are only available for 43.3\% of hotels, but \method features can be computed for 80\% of hotels --- \textbf{a relative coverage improvement of 94.8\%} ! If we combine both sources of features (review-based and \method features), we observe superior classification performance for all the \attribs.

\subsection{The effect of different groups of features}
\label{sec:eval:ablation}

Our \method method uses several groups of spatio-temporal features defined in Section~\ref{sec:method}. We now explore the relative utility of these different groups (owing to lack of space, we combined the previous-visit and next-visit features into one group of transition features). We show the ablation results in the four rightmost columns of Tables~\ref{tab:main-results-rest} and~\ref{tab:main-results-hotels}, where each column corresponds to using a different group of features.

For restaurants (Table~\ref{tab:main-results-rest}), each of the 3~temporal groups of features performs fairly well by itself, resulting in 11\%--12\% lower AUC than the full \method model. The combination of the 3~groups of temporal features yields macro-averaged AUC of 0.833 (6\% lower than \method; not shown in the table for lack of space). Interestingly, the spatial features alone (Transition) have strong predictive power, achieving AUC that is only 4\% lower than that of \method. We could not show the full results for the two individual groups of transition features owing to lack of space. However, we found the next-visit features to be slightly more predictive, as they alone achieve macro-averaged AUC of 0.829, just 6\% lower than \method; the macro-averaged AUC of the previous-visit features alone is 0.816 (8\% lower than \method).

It is insightful to consider the performance of the different feature groups for several sample \attribs. Duration features are highly informative for predicting low-price, no-reservation,
fast-food, and dessert restaurants, because customers tend to spend less time in those (for these attributes, the performance of duration features is within 5\% of \method). Similarly, both the Arrival time and Occupancy groups of features are (individually) highly predictive of breakfast, lunch, dinner, and tea \attribs, because each of these kinds of meals have characteristic hours.

Interestingly, for hotels (Table~\ref{tab:main-results-hotels}), the 3~groups of temporal features are less powerful. Individually, each group exhibits 15\%-18\% lower performance than \method, and even taken together their performance is 11\% lower. Thus, most of the \method performance is due to the spatial features (Transition), whose performance is on average within 2\% of the full \method feature set. We hypothesize this happens because people stay much longer in hotels than in restaurants (on the scale of days as opposed to hours), hence the variability in their temporal patterns is greater, and it is more difficult to find consistently predictive patterns. On the other hand, there are stronger patterns in the kinds of places that are co-visited with hotels, which explains the predictive power of the Transition features. For example, if hotel guests visit sports or recreational facilities (such as hiking trails or yoga studios), then the hotel likely appeals to fitness enthusiasts and therefore likely has a fitness center.

\subsection{The evaluation of place embeddings}
\label{sec:eval:embeddings}

In this section, we report the performance of the embeddings-based method, \methodEmb, introduced in Section~\ref{sec:method}. The results were computed using 1000 embedding dimensions (we used grid search to determine the optimal number of dimensions that maximized cross-va\-li\-da\-ted performance). Due to lack of space, we only report aggregate results across all the restaurant attributes and do not show per-\attrib results.

The classification performance of the embedding-based \methodEmb method is 6.2\% lower than the baseline, but completely eliminates the need for manual feature construction required for \method. At the same time, the improvement in classifier coverage of the \methodEmb method is 94.8\%, the same as for \methodSeq, as reported in Section~\ref{sec:eval:main}. We believe that in most practical situations, sacrificing 6.2\% in prediction accuracy is a reasonable tradeoff for doubling the coverage \underline{and} eliminating the need to craft features manually.

\section{Qualitative study}
\label{sec:qualitative}

To get a more intuitive understanding of the classification models obtained using \method, we now study the top positive and negative spatio-temporal features learned by
the trained \method model for some of the restaurant and hotel \attribs. We also plot the actual distribution of these feature values for the positive and negative classes of
the attributes (owing to lack of space, we show distribution plots for only some of the attributes).

For identifying ``romantic'' restaurants, the top positive features are those corresponding to people's visiting the restaurant on weekends, arriving at the restaurant between 8--10PM, staying for more than 90 minutes, and coming to the restaurant from a luxury hotel, museum, or performing arts theater. The top negative features correspond to people's visiting the restaurant on Mondays and Tuesdays, staying for less than 60 minutes, and arriving at the restaurant before or after visiting a grocery store or gas station. These features capture our intuition about romantic restaurants as places that people often go to for a long, leisurely meal on a weekend, often preceded by a visit to a museum, theater or concert (but presumably not often preceded by a grocery store or gas station visit).

\begin{figure}[t]
\centering
\subfigure[Duration Time Distribution]{
\includegraphics[width=0.5\textwidth]{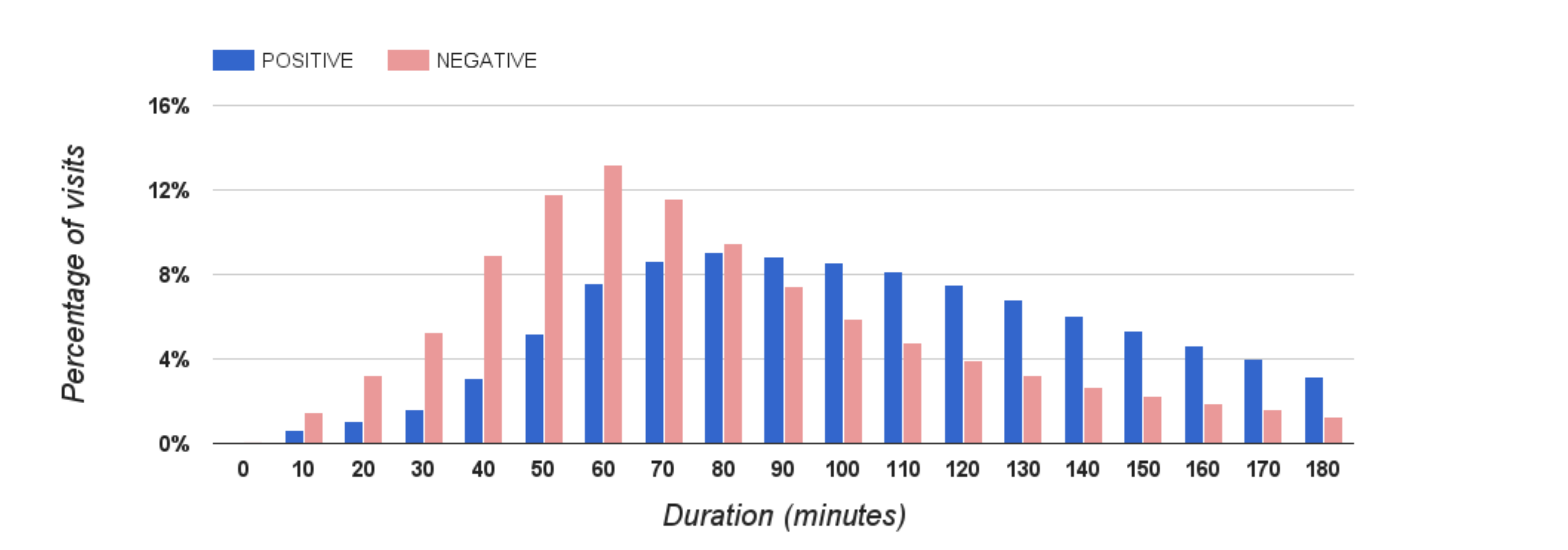}
\label{fig:romantic-duration}
}
\quad
\subfigure[Day of week distribution]{
\includegraphics[width=0.5\textwidth]{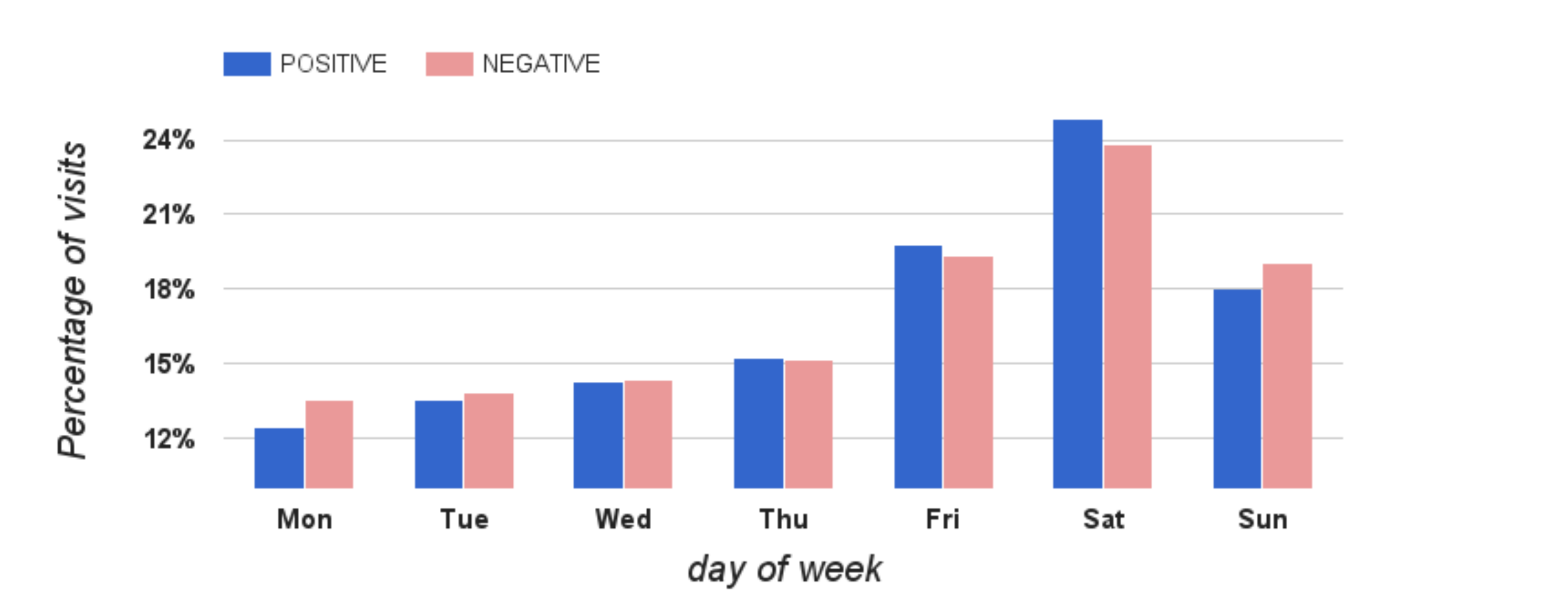}
\label{fig:romantic-day-of-week}
}
\quad
\subfigure[Previous place visit transition features]{
\includegraphics[width=0.5\textwidth]{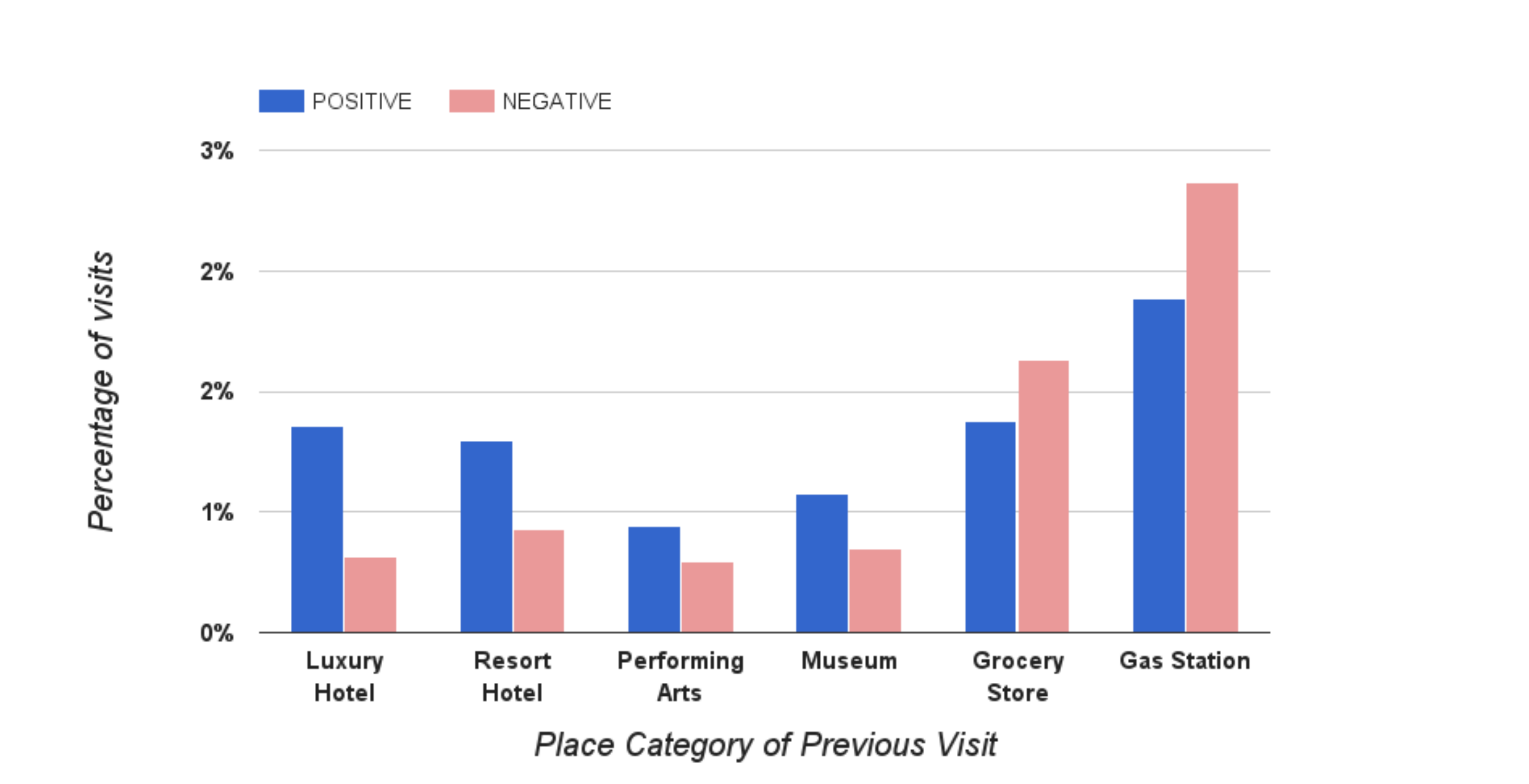}
\label{fig:romantic-transition}
}
\caption{Features for romantic restaurants.}
\label{fig:romantic-temporal}
\end{figure}

To understand these features better, in Figures \ref{fig:romantic-duration} and \ref{fig:romantic-day-of-week} we show the distributions of visit time duration and days of week of people's visits to romantic restaurants, and to restaurants not labeled as ``romantic''. The duration time distributions are clearly discriminative --- the positive class has a peak at $80$ minutes, compared to the negative class
that peaks at around $60$ minutes. The differences in the day-of-week distribution between the positive and negative classes are less pronounced, but one can still see that this difference is largest on Saturdays(positive difference) and Mondays (negative difference). These small differences in the distributions are significant enough for the model to use as discriminative features.
Figure \ref{fig:romantic-transition} plots the percentage of restaurant patrons visiting the above mentioned characteristic places before the restaurant, and confirms the utility of transition features in
the model. The figure clearly shows that patrons of romantic restaurants are much more likely to have previously visited museums, theaters and luxury hotels, compared to gas stations and supermarkets.

\begin{figure}[t]
\centering
\subfigure[Arrival Time Distribution]{
\includegraphics[width=0.5\textwidth]{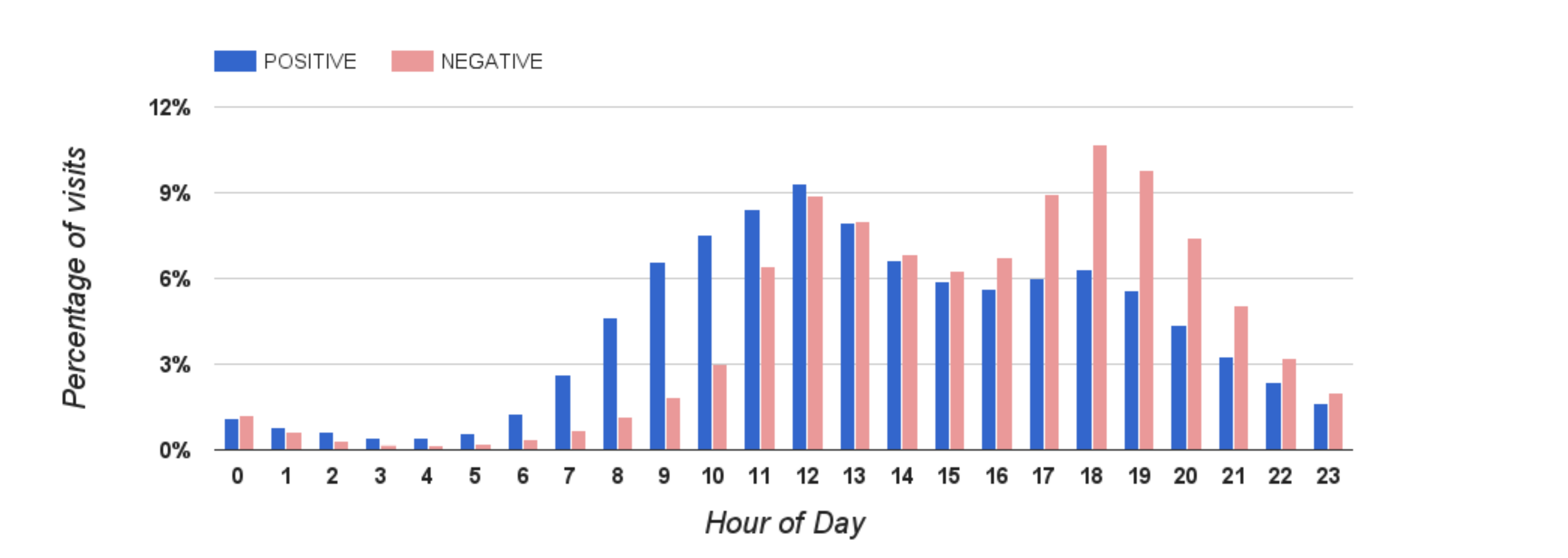}
\label{fig:breakfast-arrival}
}
\quad
\subfigure[Occupancy distribution]{
\includegraphics[width=0.5\textwidth]{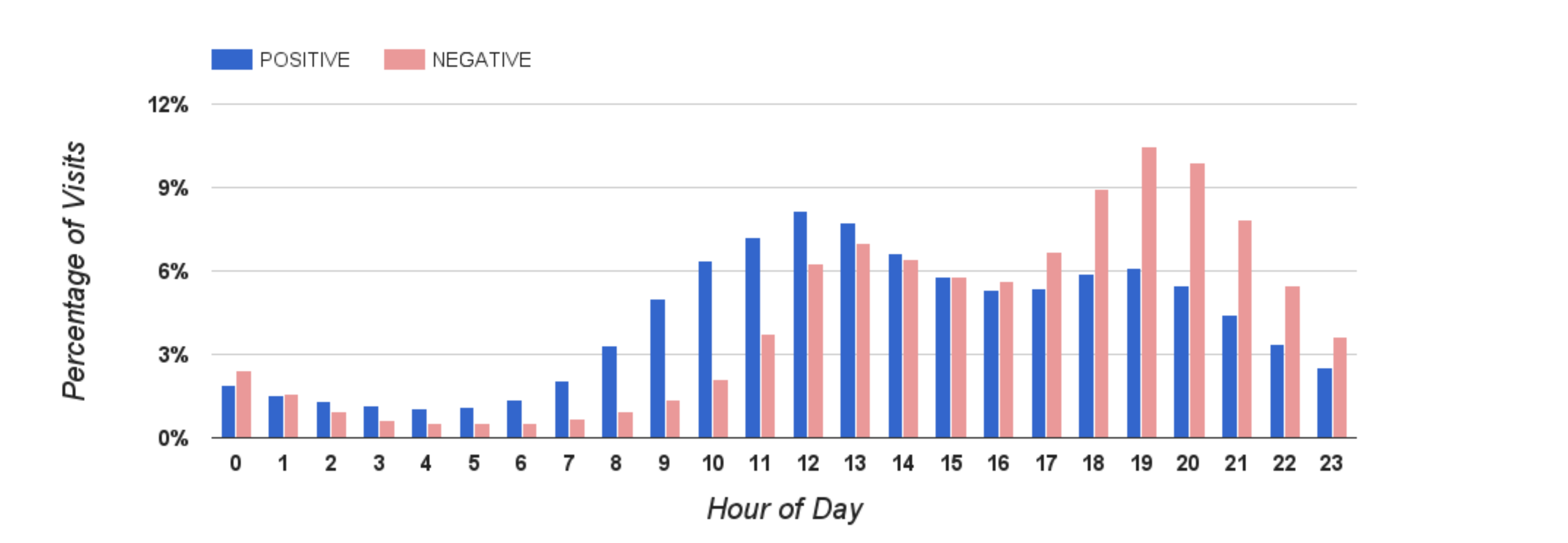}
\label{fig:breakfast-occupancy}
}
\caption{Features for restaurants offering breakfast.}
\label{fig:breakfast-temporal}
\end{figure}

For identifying restaurants offering ``breakfast'', temporal features are, not surprisingly, the most informative features in the classification model. The top positive features include arriving at the restaurant between 8--11AM, and staying at the restaurant (occupancy) from 9--11AM. The top negative features include arrival times between 12--2PM, indicating that the restaurant is more likely to be popular for lunch than for breakfast. Figures \ref{fig:breakfast-arrival} and \ref{fig:breakfast-occupancy} plot the distribution of these features for the ``breakfast'' attribute, and highlight the slightly earlier arrival and occupancy times for the positive 
class that are picked up by the model.

For identifying restaurants or cafes that offer WiFi, the top positive features correspond to people staying at the place between 4--7PM or between 10--11PM, visiting the place on weekdays, and visit duration times between 30 to 80 minutes. Other positive features include visiting a work location shortly before or after the cafe. Indeed, these features match our expectations of cafes that offer WiFi --- they are usually busy late afternoons and late evenings on weekdays, and are frequented by people who use WiFi to work and therefore stay longer at the place.
The top negative features include those corresponding to people's visiting the place on Fridays and Saturdays, and staying at the place for less than 30 minutes.

For predicting restaurants that offer ``takeout'', the top positive features are again related to duration --- a large fraction of people stay at such places for less than 20 minutes. However, some of the top 
features are not so obvious, such as visiting a work location immediately before or after the restaurant (positive feature), and visiting a shopping center or bar after the restaurant (negative feature).

Another \attrib where the predictive power of \method is, at first glance, rather unexpected is ``healthy'' restaurants.
The top positive features
correspond to visiting the restaurant in the middle of the week (Tuesdays, Wednesdays and Thursdays) and going to work immediately after the restaurant. The top negative features
correspond to visits on weekends, and visiting a bar immediately before or after the restaurant. This hints at some interesting socio-cultural trends; for example, people are more likely to frequent healthier restaurants in the middle of the week than during the weekend, and people visiting healthy restaurants are less likely to visit bars before or afterwards.

\begin{figure}[t]
\centering
\includegraphics[width=0.5\textwidth]{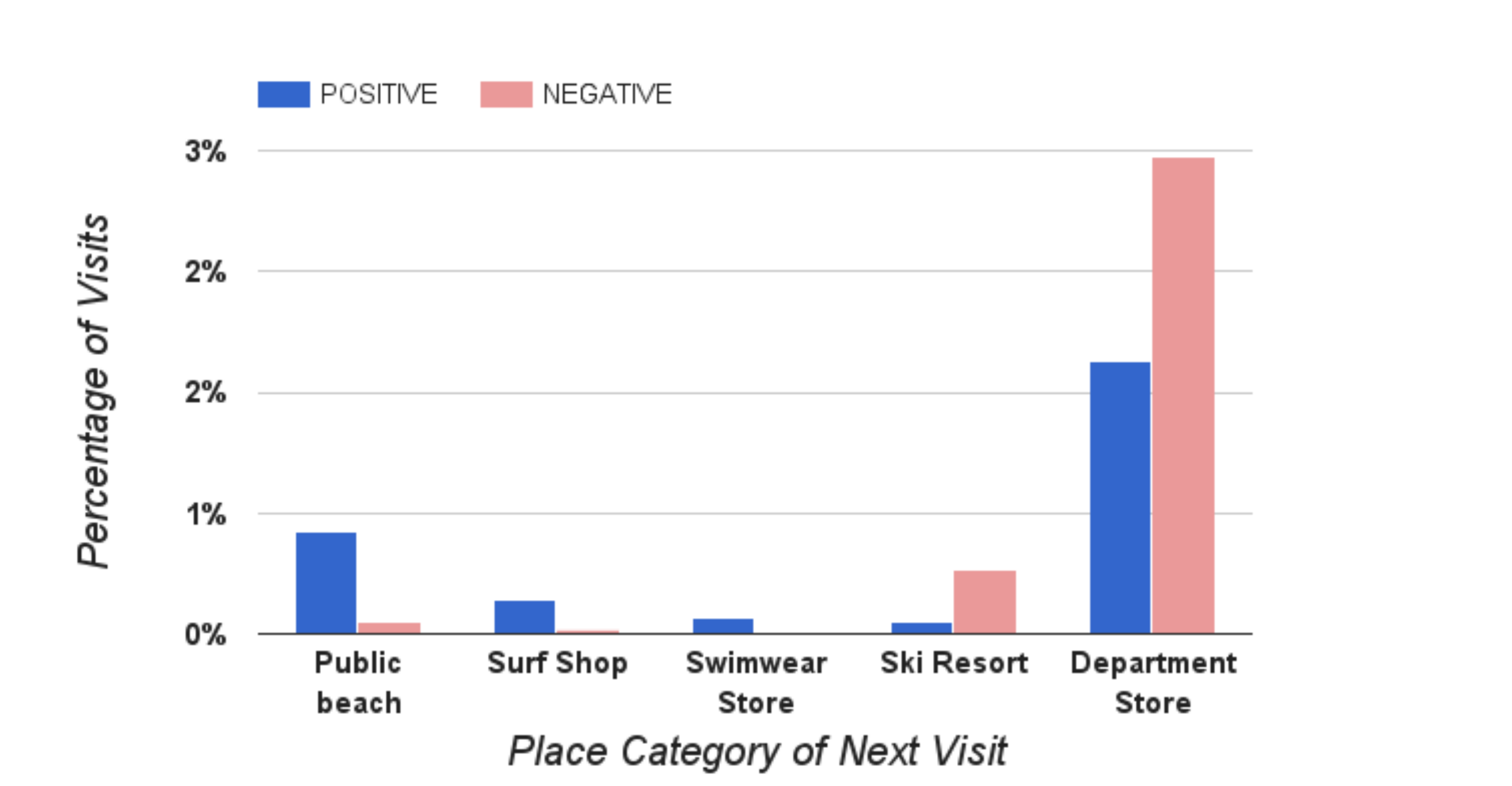}
\caption{Transition features for hotel beach access}
\label{fig:hotel-transition}
\end{figure}

We also observe interesting trends when studying the top positive and negative features for some of the hotel \attribs. In contrast to predicting restaurant attributes, duration and occupancy features are less important here than transition features, which capture the types of places visited by the hotel guests. For example, to identify hotels with ``beach access'', the top features correspond to visits to surf shops, swimwear stores and beaches, while the top negative features include visits to water parks and ski resorts. Figure \ref{fig:hotel-transition} shows the proportion of hotel guests visiting these places from the hotel. On average, guests at hotels with beach-access are
$10$ times more likely to visit surf shops, swimwear stores and beaches, $3$ times less likely to visit ski resorts, and $2$ times less likely to visit department stores.

\section{Conclusions}
\label{sec:discussion}

Many online services recommend brick-and-mortar businesses such as restaurants or hotels based on their \attribs. These \attribs are often difficult to obtain, and previous approaches based on crowdsourcing and mining review text  were challenging to scale, and hence had limited coverage. We presented an approach to \emph{predict} numerous place \attribs using spatio-temporal features, which characterize how large populations of people go in and out of these places. Our method, \method, uses several groups of spatial and temporal features, and its variant, \methodEmb, uses embeddings to completely eliminate the need for manual feature construction. The key idea is to derive signals from anonymous aggregated observational data. This allows us to reliably predict dozens of \attribs of businesses without ever visiting them or talking to people who did.

In an extensive empirical evaluation, we compared our methods to a baseline that uses web reviews for restaurants and hotels. Our \method method was able to reliably classify numerous place \attribs while nearly doubling the coverage of the baseline (review-based) classifier, and offered comparable or superior performance. In our future work, we plan to explore joint modeling of attributes, as well as experiment with cross-products of \method features.


\section*{Acknowledgments}

We would like to thank John Giannandrea and Chandu Thota for making this work possible. We would also like to thank Ravi Kumar, Bo Pang, Finnegan Southey, Mukund Sundararajan, and Andrew Tomkins for stimulating discussions that helped greatly improve this paper.

\bibliographystyle{acm}
\bibliography{place-attrib}

\end{document}